\documentclass{preaa}  
\usepackage{hyperref}
\usepackage{graphicx}
\usepackage{txfonts}
\usepackage{subcaption}

\begin{document}

   \title{The data center for the Spectrometer and Telescope for Imaging X-rays (STIX) onboard Solar Orbiter}

   \subtitle{}

   \author{Hualin Xiao
          \inst{1}
          \and 
          Shane Maloney 
          \inst{2}
          \and S\"am Krucker\inst{1}
          \and Ewan Dickson \inst{3}
          \and Paolo Massa \inst{4}
          \and Erica Lastufka \inst{1}
          \and Andrea Francesco Battaglia\inst{1,5}
          \and László Etesi \inst{1}
          \and Nicky Hochmuth \inst{6}
          \and Fr\'ed\'eric Schuller \inst{7}
          \and Daniel F.\ Ryan \inst{1}
          \and Olivier Limousin \inst{8}
          \and Hannah Collier \inst{1,5}
          \and Alexander Warmuth \inst{7}
          \and Michele Piana \inst{9,10}
         }

   \institute{University of Applied Sciences and Arts Northwestern Switzerland (FHNW), 5200 Windisch, Switzerland \\
              \email{hualin.xiao@fhnw.ch}
         \and           Astronomy and Astrophysics Section, School of Cosmic Physics, Dublin Institute of Advanced Studies, 31 Fitzwilliam Place Dublin 2, D02XF86, Ireland
         \and University of Graz, Universitätspl. 3, 8010 Graz, Austria
      \and Department of Physics \& Astronomy, Western Kentucky University, Bowling Green, KY 42101, USA
          \and ETH Z\"urich, R\"amistrasse 101, 8092 Z\"urich, Switzerland
         \and Ateleris, Badenerstrasse 13, CH-5200 Brugg Switzerland
         \and Leibniz-Institut für Astrophysik Potsdam (AIP), An der Sternwarte 16, D-14482 Potsdam, Germany
         \and Université Paris-Saclay, Université Paris Cité, CEA, CNRS, AIM, 91191 Gif-sur-Yvette, France
         \and 
         MIDA, Dipartimento di Matematica, Università di Genova, via Dodecaneso 35, I-16146 Genova, Italy
         \and 
            Istituto Nazionale di Astrofisica, Osservatorio Astrofisico di Torino, Via Osservatorio 20, I-10025 Pino Torinese, Italy
             }

   \date{\today}

 
  \abstract
   {
   The Spectrometer and Telescope for Imaging X-rays (STIX) on board Solar Orbiter observes solar X-ray emission in the range of 4\,--\,150\,keV and produces spectra and images of solar flares over a wide range of flare magnitudes.  During nominal operation, STIX continuously generates data. 
   A constant data flow requires fully automated data-processing pipelines to process and analyze the data, and a data platform to manage, visualize, and distribute the data products to the scientific community.  
   } 
   {The STIX Data Center has been built to fulfill these needs.  In this paper, we outline its main components to help the community better understand the tools and data it provides.
   }
   {
   The STIX Data Center is operated at the University of Applied Sciences and Arts Northwestern Switzerland (FHNW) and consists of automated processing pipelines and a data platform.  The pipelines process STIX telemetry data, perform common analysis tasks, and generate data products at different processing levels. They have been designed to operate fully automatically with minimal human intervention.  The data platform provides web-based user interfaces and application programmable interfaces for searching and downloading STIX data products.
   }
   {
   The STIX Data Center has been operating successfully for more than two years. The platform facilitates instrument operations and provides vital support to STIX data users.}
 {}
 \titlerunning{STIX Data Center}
   \maketitle


\section{Introduction}
Solar Orbiter is an ESA-led mission with a suite of ten remote-sensing and in situ instruments to study the interaction between the Sun and the heliosphere. Launched on 10 February 2020 into a heliocentric orbit, its nominal mission will last seven years with a planned extension of three additional years \citep{Mueller2020}.
Solar Orbiter's closest approach to the Sun is 0.28\,AU and its orbit will shift out of the ecliptic as the mission progresses, providing better views of the solar poles \citep{SolarOrbiter2020}.

Solar Orbiter's Spectrometer Telescope for Imaging X-rays \citep[STIX;][]{stix2020} observes X-rays from 4 to 150 keV.  Its main scientific objective is to study the hot solar plasma and accelerated electrons produced during solar flares.
STIX consists of 32 subcollimators, each with a pair of slightly offset grids that encode spatial information in a Moiré pattern which is detected by a pixelated cadmium telluride (CdTe) detector. Images with an angular resolution of a few arcseconds can be produced on the ground with Fourier-based image reconstruction techniques.  STIX's detectors are spectroscopic and so they enable the thermal plasma and accelerated electrons to be characterized via the spectra they produce.
They also enable images to be produced in any energy range between 4 to 150 keV and over any integration time greater than $\sim$0.5\,s.
This allows scientists to customize the imaging based on their science goals and observational constraints. Hence, STIX provides observations of the intensity, energy, dynamics, and location of accelerated electrons and heated plasma during solar flares. For more information on STIX instrumentation and its scientific capabilities, readers can refer to \citet{stix2020}.

During normal operations, STIX observes continuously. HouseKeeping (HK) and QuickLook (QL) data are generated automatically and sent to Earth at regular intervals.
Science-quality data products are generated on board only in response to requests from the ground.
These products are clipped and summed in time, energy, and pixels depending on the quality of the observation and then further compressed to optimize the telemetry needs on the instrument.
This results in hundreds of different types of raw telemetry data packets.
To handle these, the STIX Data Center has developed a data platform and automated data-processing pipelines which assist with instrument operations, simplify STIX data analysis, and increase data accessibility for the solar physics community.
The pipelines convert raw telemetry data into QLs and data products at different processing levels which can be used for scientific analysis. The platform stores all STIX data products and provides web-based tools for users to explore and analyze STIX data.

 In this paper, we describe the processing pipelines, core analysis algorithms, data products, and other tools used and provided by the STIX Data Center.  The aim is to help users better understand the data center and hence how to best utilize it and the STIX instrument.

\section{STIX telemetry data products}
\label{sec:raw-data}
STIX generates many different types of raw telemetry packets.
However, only three categories are relevant to scientific users: HK, QL, and  science data. HK and QL data are directed to the low-latency data stored in the spacecraft's solid-state mass memory (SSMM) and sent to the ground with the highest priority.
Except for the raw pixel data product, all count- and trigger-based values in QL and  science data products are compressed with an integer compression algorithm \citep{stix2020}.

\subsection{HK  data}
 The HK data monitor the instrument's status and performance in order to ensure it functions properly. 
 STIX generates HK packets continuously while being powered.  
 The HK packets include details such as the instrument's temperature, 
 voltages, currents, CPU usage, the status of the attenuator switches, 
 trigger rates, file system information, and aspect system readouts \citep{Warmuth2020, stix2020}. STIX typically generates a  HK packet every 64 seconds. 
 
\subsection{QL data}
The QL data are only generated when STIX is in the nominal observation mode, as they require the detectors to be powered and operating. There are five types of QL data:
\begin{itemize}
\item {\bf QL light curves} contain time series of 4-second-integrated,  detector-summed counts in five energy bands.  Furthermore, they include the corresponding detector-summed triggers, and the rate control regime state \citep[i.e., attenutator state; see ][]{stix2020}.  We note that the QL light curves do not include time series from STIX's two special detectors, the BacKGround monitor (BKG) and the coarse flare locator (CFL). 
It is worth mentioning that STIX does not correct for detector dead time, absorption of X-rays by the entrance and grids,
impacts of the presence of the attenuator,  or rate control regime states on board.
Hence the effects of these phenomena are present in the QL light curves.

\item {\bf QL background light curves}: STIX uses the background monitor -- a detector with a special grid aperture -- to monitor both the X-ray background and the intense unattenuated X-ray fluxes from large solar flares. It consists of an open front grid window and a rear grid window that is fully opaque except for six small openings. QuickLook background light curves contain counts and triggers of all the detector pixels integrated over 8 seconds in the same five energy bands as the QL light curves. 

\item {\bf QL variance data} are the onboard computed variance of 40 successive 
detector-summed count rates based on a 0.1-second integration.

\item {\bf QL spectra} are produced for each detector by integrating over all its pixels and accumulated over 32 seconds. During nominal operations, a QL spectrum for each detector is generated every 1024 seconds.

\item {\bf Calibration spectra} are high energy resolution, per pixel, energy spectra in ADC units (ADU) accumulated for photons emitted by $^{133}$Ba radioactive sources located within the STIX instrument. 
STIX generates a calibration spectrum for each pixel every 24 hours during normal operations. These are used to calibrate the gain of each pixel on the ground by observing the locations of the well-known spectral line peaks. 
If the gain is seen to shift, a new ADU to keV conversion table is uploaded for each pixel from the ground. 
\end{itemize}

\subsection{Science data }
Science data products are only generated and down-linked in response to a data request from the ground. They do not present a continuous record of solar activity because of the limited bandwidth and onboard storage.  Instead, they cover scientifically interesting periods identified on the ground from the low-telemetry QL data. Science data are generated from the pixel data stored in the STIX onboard archive memory.  They are rebinned in time, energy, pixels, and detectors in accordance with the parameters of the data request.  These parameters are chosen on the ground on a case-by-case basis to optimize both telemetry efficiency and scientific return.  Due to the limited storage available on board, raw pixel data are overwritten after a few months.  Therefore, bulk science data (BSD) requests must be made within the following weeks of observations being made.

STIX can generate six different types of  science data products: 
\begin{itemize}
 \item {\bf Raw pixel data} are the least processed data product, which provides time-binned pixel counts in the highest available time and energy resolution, generally with a dynamic accumulation time between 0.5~s and 20~s depending on solar activity. The raw pixel values are not integer-compressed, and hence require a large amount of telemetry.  Therefore, they are rarely requested and are used primarily for testing and verification.
 
\item {\bf Compressed pixel data} are based on the raw pixel data product.  They combine the highly resolved pixel counts into larger user-defined time and energy bins. In addition to the rebinning, they are integer-compressed on board before being sent to the ground. This is the most common BSD product and is well suited for scientific analysis. 

\item {\bf Summed and compressed pixel data} are also based on the raw pixel data product.  Similar to the compressed pixel data, they combine the raw pixel counts into larger user-defined time and energy bins, but they also sum the counts over multiple pixels. The pixel sums can be configured by telecommand. By default, the top and bottom pixels in the same column are summed, reducing the apparent number of pixels per detector to four\citep{stix2020}. Therefore, it is a suitable option when telemetry is limited.  In addition, the summed pixel data values are integer-compressed.

\item {\bf Visibility data}: Each STIX subcollimator measures a spatial Fourier component on the sky which can be combined into images on the ground using indirect imaging algorithms.  Each Fourier component is encoded into an intensity sine wave which is sampled at four locations by the subcollimator's pixel columns.  This information can be further encoded into a complex number which gives the amplitude and phase of the sine wave \citep{paolo2022}.  This format, known as visibility, is twice as data efficient as the summed pixel column values.  Because STIX has been given a larger telemetry budget than originally anticipated, this data product is rarely requested because the visibility calculation can be done slightly more accurately on the ground.  However, should telemetry be significantly reduced in the future, this product can be requested with a limited loss of scientific value.

\item {\bf Spectrogram data} are summed over selected pixels, detectors, and time-energy bins. The pixel and detector summing does not lead to a significant loss of spectral resolution, but it does eliminate the spatial information.  In almost all cases, however, spectrogram data are equally useful for spectral analysis as compressed pixel data, but they require two orders of magnitude less telemetry.  This means that spectral analysis can be performed at high resolution and time cadence for longer time periods than would otherwise be possible.  Hence, along with compressed pixel data, spectrogram data are the most requested science data product. It is worth mentioning that the spectrogram data are available at the highest resolution possible at all times that STIX is in the nominal observation mode since January 2022. 

\item {\bf High time-resolution aspect data}: STIX stores readouts from the four photodiodes in the aspect system in the onboard archive memory at a resolution of 16~ms. So-called burst-mode aspect data can be later requested at full resolution, or binned to user-specified resolution. This is especially useful to accurately measure the direction where STIX is pointing, even when the spacecraft attitude is changing rapidly.
\end{itemize}
Science data are transmitted to the ground in raw binary packets.  
These are labeled level 0 at the STIX Data Center.  

\section{Data flow}
\begin{figure*}[ht]
    \centering
    \includegraphics[width=0.9\linewidth]{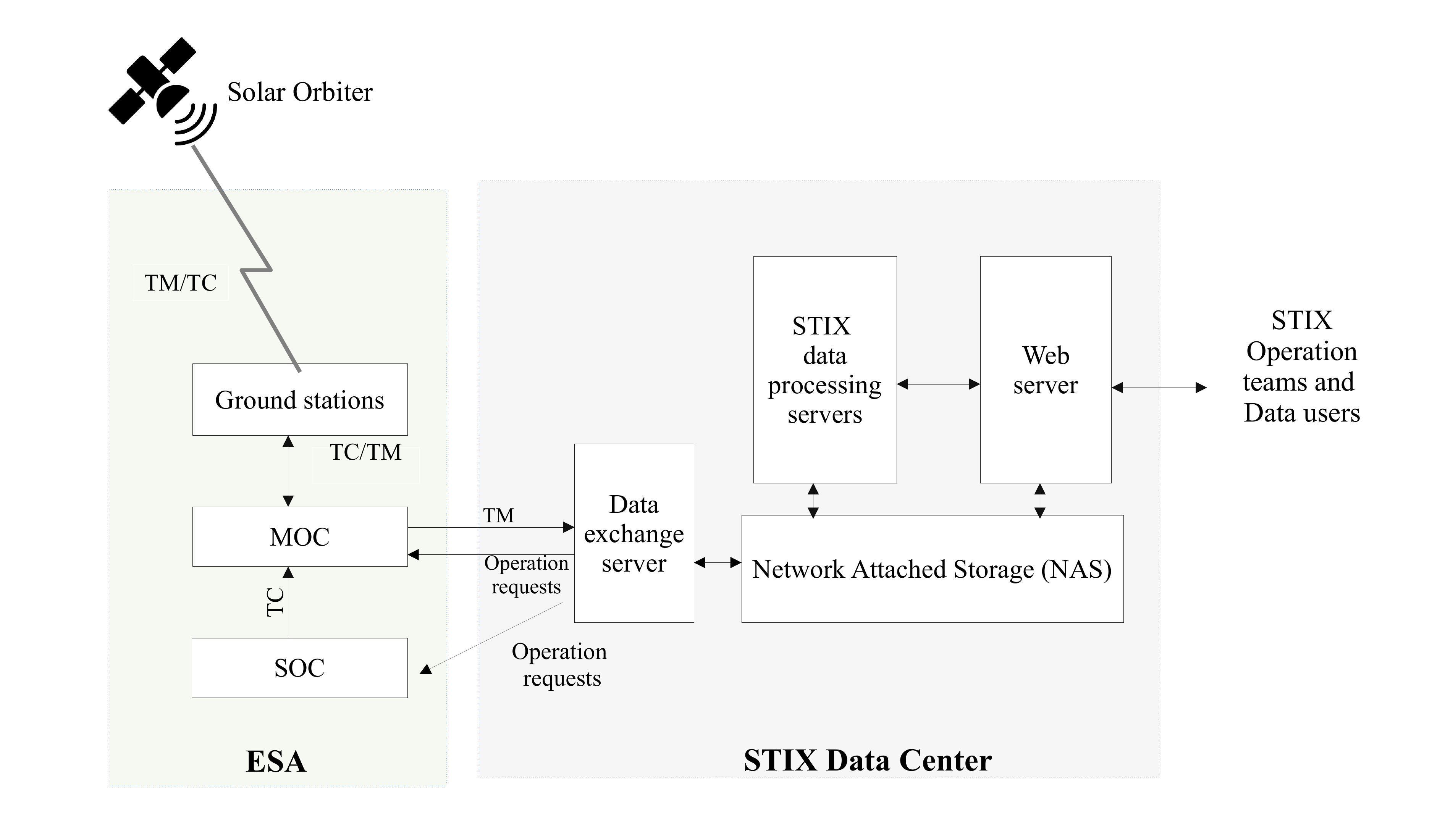}
    \caption{ STIX Data Center hardware architecture and  data flow. Solar Orbiter telemetry data are processed at the Mission Operations Center (MOC) and delivered to the instrument team via the generic file transfer system (GFTS). 
    The STIX Data Center currently utilizes five servers: one for data exchange, two for running the data-processing pipelines, one for the web server, and one for data storage (network attached storage or NAS). STIX data products are stored on the NAS server. The data exchange server is also used to deliver instrument operations requests (IORs) to the MOC, and payload direct operation requests (PDORs)  to  the Science Operations Center (SOC). 
    }
    \label{fig:data-flow}
\end{figure*}
During nominal science operations, low-latency data are transmitted to Earth during every ground station pass regardless of orbital geometry, whereas science data are only downlinked when the bandwidth permits. Every instrument team is responsible for keeping the transmitted data volume within preallocated boundaries.

Telemetry data received by ground stations are processed by the Mission Control System at the Solar Orbiter Mission Operations Center (MOC). Telemetry data after undergoing quality checks, packet extraction, and filling are stored in the  EGOS Data Dissemination System (EDDS) \citep{egos,EDDS}. They are distributed to the instrument teams regularly via the generic file transfer system (GFTS).  
As shown in Fig.~\ref{fig:data-flow},  the STIX Data Center uses a dedicated  server to exchange data between the STIX Data Center and ESA.  In addition to receiving telemetry, it is also used to deliver operation requests   to the Science Operations Center (SOC) or the MOC.  
Telemetry files received by the data exchange server are transferred to network attached storage (NAS), and are subsequently processed by the pipelines introduced in the next chapter.

Low-latency telemetry data generally arrive at the STIX Data Center within a few days, depending on the next ground station pass. However, science data may be delayed 
several weeks after being generated on  board due to possibly low bandwidth or priority.
In addition to telemetry data, the STIX Data Center receives auxiliary data \citep{spice1996,spice2018} from the SOC, which contain information on spacecraft ephemeris, attitude, and calibration factors required for the conversion of onboard timestamps to UTC times. 

\section{Data-processing pipelines}
\subsection{Data-processing pipeline overview}

\begin{figure*}[htb]
    \centering
    \includegraphics[width=0.9\linewidth]{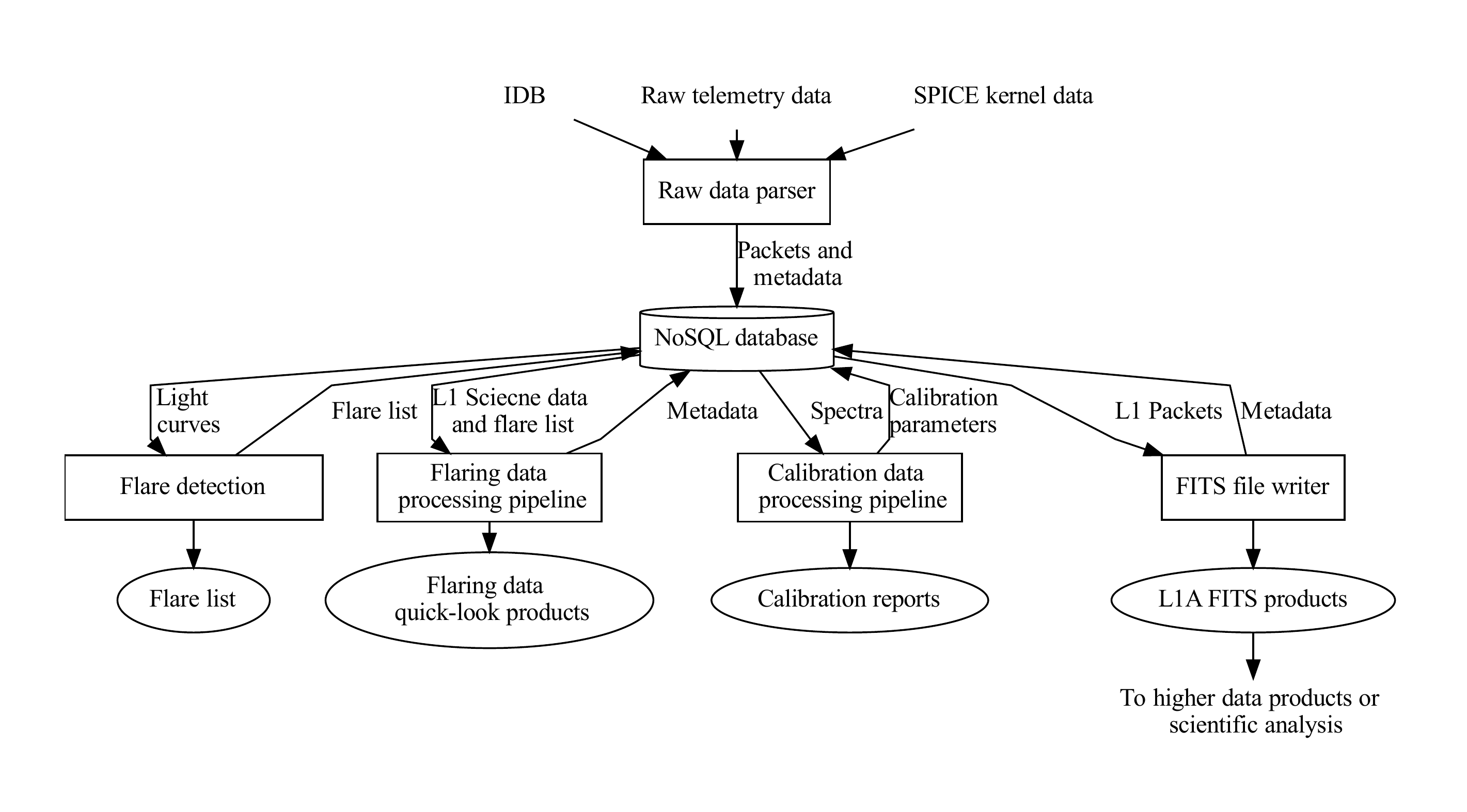}
    \caption{Telemetry processing pipelines at the STIX Data Center. Raw packets are processed to
level 1, which includes parsing, integer decompressing, and timestamp conversion. Level-1 packets are recorded in a NoSQL database. Then they are selected and processed in four different paths, each of which is responsible for processing specific data types.    }
    \label{fig:main_pipelines}
\end{figure*}
STIX telemetry data are processed immediately upon reception at the STIX Data Center by the data-processing pipeline  
shown in Fig.~\ref{fig:main_pipelines}.  Raw packets are processed to
level 1 (L1) which includes parsing, integer decompressing, and timestamp conversion. 
Level-1 packets are stored in a tree-like structure and recorded in a NoSQL database. Then they are selected and processed in four different paths. 
In the first, the HK, QL, and science data are used to create L1 FITS files. 
The second path computes energy calibration factors from calibration spectra.   
In the third path, solar flares are identified by applying a detection algorithm to QL data. 
The fourth path performs common analyses of science data.

\subsection{Raw packet parsing}
Raw telemetry data at the STIX Data Center are stored as binary packets. 
Each packet contains a fixed-length header and a list of parameters that vary with the type of packet.  The parsing of parameters is based on information in the Mission Information Base (MIB), which contains the name and length of each parameter for each type of packet. 
The parsed packets contain raw values of parameters, which need to be converted to physical values. 
Raw values of spacecraft-local times are converted to UTC times with the latest version of SPICE kernels \citep{spice1996,spice2018,spicedoi}.  Raw values of HK parameters are converted to physical values using the ground-calibrated conversion factors stored in the MIB. 
Compressed counts in science data are decompressed using a lookup table. 
After the above processing steps, packets are organized in a tree-like structure. 
They are considered L1 packets and recorded in a collection in the NoSQL database. 
The NoSQL database is schemaless, which means that the data format of each record can be different and a predefinition of the data format is not required.
This makes it very suitable for storing the tree-like structure L1 data packets.  
The NoSQL database is extremely convenient  for tasks such as searching, sorting, organizing, merging data packets, and verifying data integrity.

In addition to L1 packets, other metadata, such as filenames, SPICE kernel version, and the MIB version, are recorded in another collection in the NoSQL database.
This allows for a rapid query of the associated raw telemetry data.

\subsubsection{FITS products}
The flexible image transport system (FITS) is a portable file standard widely used in astronomy to store, transmit, and manipulate scientific images, tables, and associated data \citep{fits}.
Therefore, the FITS format is adopted by the STIX Data Center to store the standard data products. 
After parsing each new raw telemetry file, HK, QL, and science packets are sequentially selected from the NoSQL database and merged after passing checks for data integrity and consistency.  The merged data as well as the associated metadata and auxiliary data are written to FITS files.  In the meantime, their metadata are recorded in a collection in NoSQL, which allows for the fast query of the products. The FITS files, created from L1 packets immediately upon the parsing process, are defined as level-1A (prereleased L1) products.  They are used by some subsequent data-processing pipelines.  

FITS files are recreated 
in a similar pipeline after a few days to weeks after all inputs are validated manually. The created FITS files are regarded as the formal L1 products at the STIX Data Center. 
In most cases, the FITS files at level-1A and L1 are almost identical, except that the level-1A FITS files may use predicted ephemeris data. In addition, level-1A FITS files may not contain all header keywords required by the ESA standard.  As such, L1 FITS files are recommended for STIX data users whenever they are available. 

\subsection{Energy calibration}
\begin{figure}
 \centering
  \includegraphics[width=0.8\linewidth]{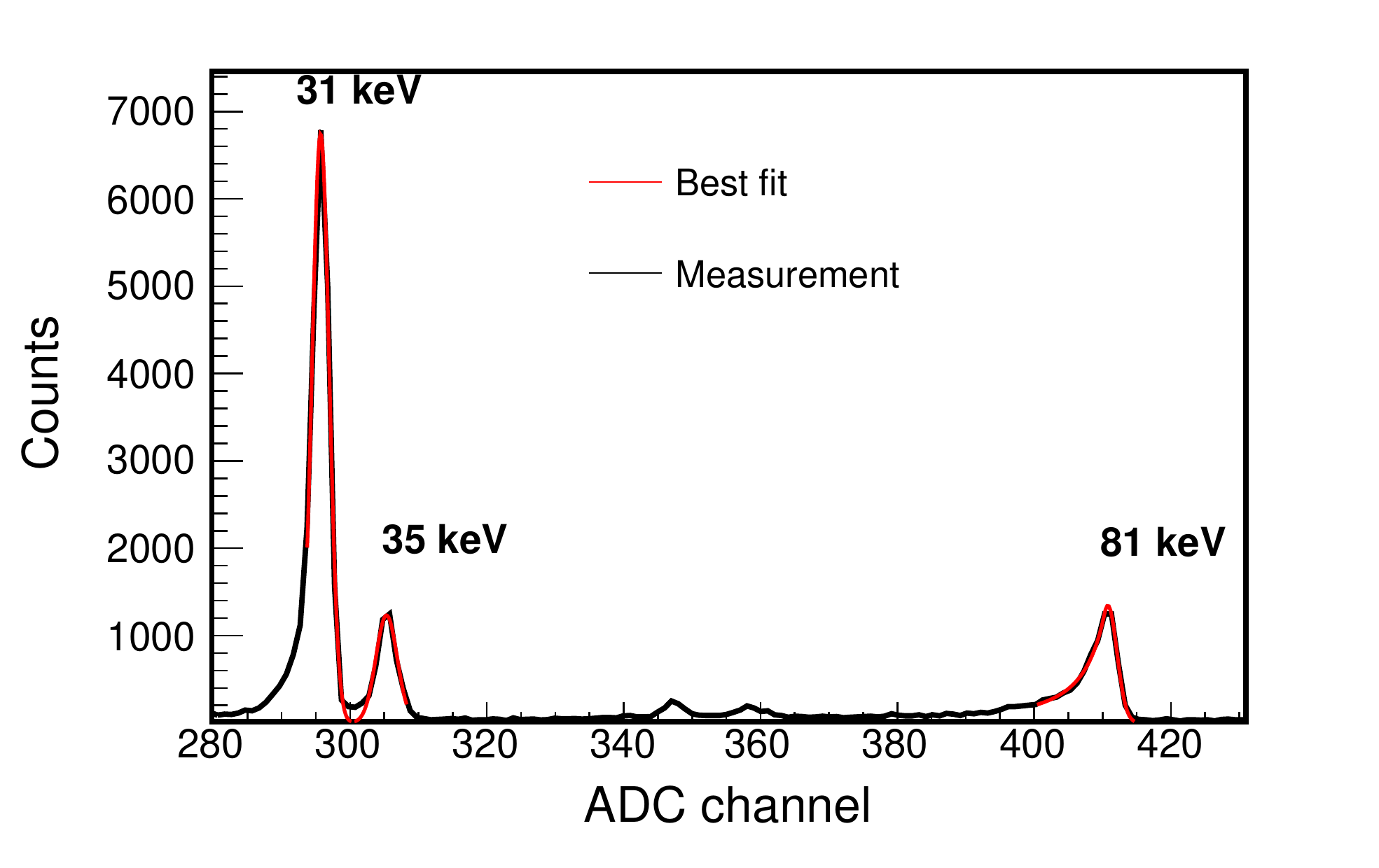}
  \caption{Example of  STIX in-flight calibration spectrum recorded by a single pixel.
  The most prominent peaks, from left to right, are the 31, 35, and 81 keV photopeaks. The first two are fitted with a double-Gaussian function, and the high energy peak with a crystal-ball function. }
    \label{fig:cal-fit}
\end{figure}
STIX converts ADC channels to "science energy channels" on board by the FPGA using an energy lookup table (ELUT), 
which defines the ADC channel edges for each science channel for each pixel.
An ELUT can be constructed using energy conversion factors determined from calibration runs. 
STIX continuously accumulates an energy spectrum (in ADC units) for each pixel separately for events from the onboard Ba$^{133}$ sources and formats a spectrum typically every 24 hours. 
Fig.~\ref{fig:cal-fit} shows an example of such a spectrum.  The three most prominent peaks are produced by photons of 31 keV, 35 keV, and 81 keV energy from the calibration sources. To determine the positions of the photopeaks, the first two peaks are fitted with a double-Gaussian function, and the third peak with a crystal-ball function \citep{crsystallball},
which consists of a Gaussian core 
and a power-law low-end tail, below a certain threshold.
Then a linear fit is made to the positions and keV energies whose slope and intercept give the gain (i.e.,\ the ADC to energy conversion factor) and baseline, respectively.
The ECC method \cite[see ][]{ecc,ecc2} is another method often used by the STIX team to determine the calibration factors.  We found that the results of the two methods are consistent within 1$\sigma$.

The above steps are performed for each calibration spectrum once the data are available at the STIX Data Center. 
The calibration factors are recorded in a collection in the NoSQL database and used for further correction of energy bins in offline data analysis.   
Once significant changes in the calibration factors are observed, the STIX operations team creates a new ELUT and uploads it to STIX.

\subsection{Solar flare identification}
STIX identifies solar flares on board based on detector count rates, and the results are included in the QL data \citep{stix2020}.  However, the data only provide limited information on flares due to the constraints of the telemetry budget and onboard computing resources.  Moreover, microflares are not reported due to the relatively high trigger threshold.
It is necessary to maintain a flare list on the ground for requesting science data and also for users to find events of interest. 

\begin{figure}
  \centering
  \includegraphics[width=0.8\linewidth]{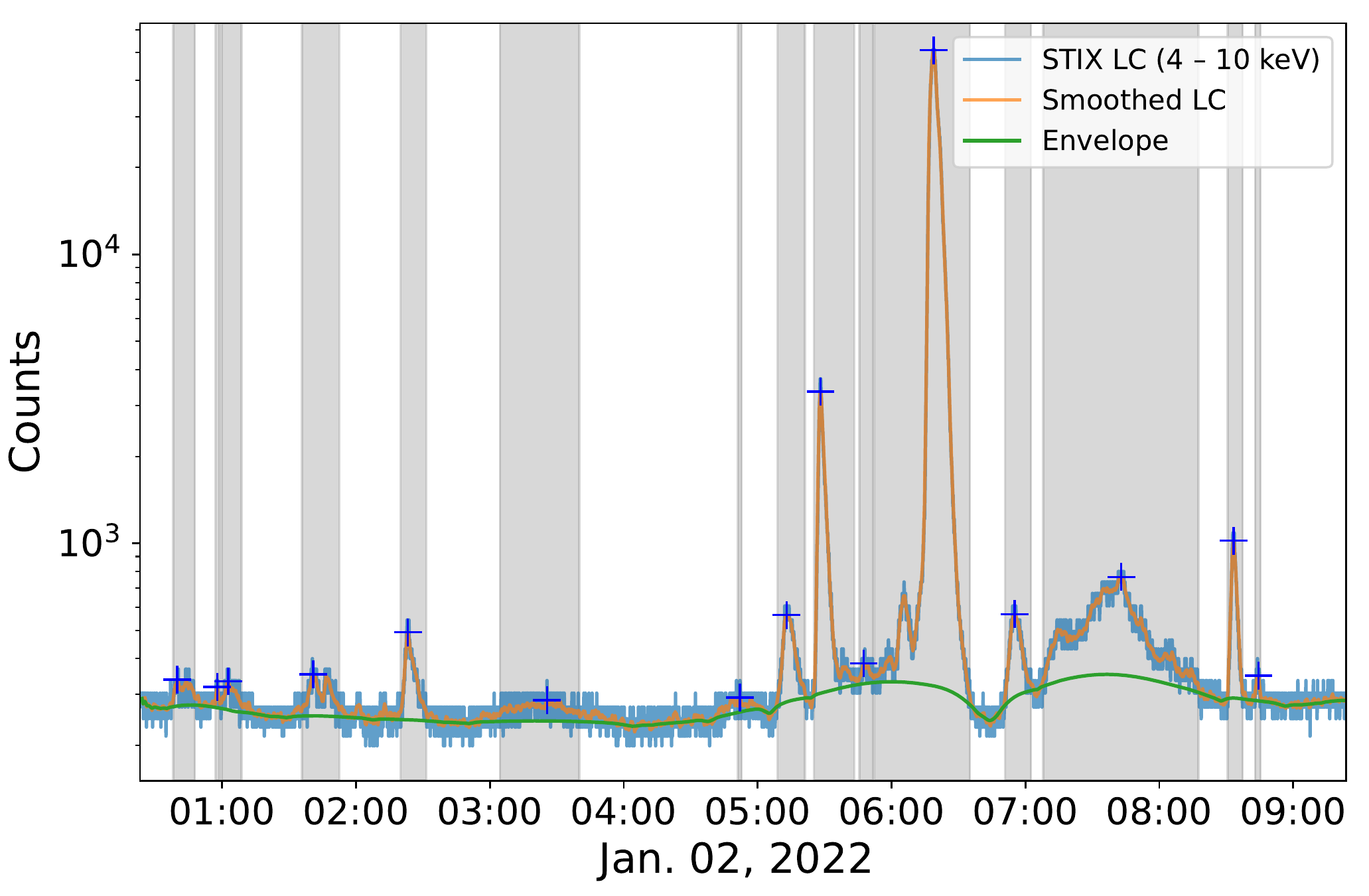} 
  \caption{STIX 4\,--\,10\,keV QL light curve for 2022-01-02T00:30:00 -- 2022-01-02T09:30:00 (UTC) and  identified flares. The orange curve is the smoothed light curve using a moving average filter with a time window of 1 minute. The green curve is the estimated signal envelope using the SNIP algorithm.  The identified peaks are marked with plus signs, and the gray ranges show their time ranges.
  }
  \label{fig:flare-det}
\end{figure}
Using QL light curves, solar flares can be identified in greater detail on the ground. 
The ground identification procedure includes the following steps:
\begin{itemize}
  \item Light curve smoothing: The selected light curve is filtered using an unweighted moving average filter with a time window of 1 min. This can smooth out statistical fluctuations and electric surge spikes.
  \item Envelope subtraction: A flare may last hours, and there may be short-duration pulses lying on the envelope (the main pulse) in the light curve.  To facilitate the identification of those short-duration pulses, the envelope is subtracted from the smoothed light curve, which is estimated using the SNIP algorithm \citep{snip}. 
  \item Identification of flare peaks: We consider that a flare is detected if the peak count rate after envelope subtraction lies beyond two standard deviations of the mean count rate during quiet Sun periods. The flare start and stop times are given at the times exceeding this threshold.  
  \item Merging of flare peaks: Two peaks are considered from one single flare if the peak times differ by less than 5 min. This can reduce the number of reported flares and simplify data analysis. We note that this can also affect any statistical flare distribution derived from this list.
\end{itemize}

As an example, Fig.~\ref{fig:flare-det} shows the STIX QL light curve in the energy range 4 -- 10 keV, recorded from  2022-01-02T00:30:00 to 2022-01-02T09:30:00 (UTC).  The orange curve is the smoothed light curve.  The identified peaks are marked with the plus signs, and the colored ranges show the time ranges.

The above steps are repeated for QL light curves of the other four higher energies, which provide information on the upper limit of the X-ray energy of the flare. The time ranges, peak count rate, and total counts as well as the corresponding ephemeris data of the identified flares are  stored in a collection called the flare list in the NoSQL database, which is later used for the creation of data requests (see \ref{sec:datareq}).

\subsection{Solar flare standard analysis pipeline}
\subsubsection{Estimation of solar flare GOES class }
\begin{figure}
  \centering
  \includegraphics[width=0.8\linewidth]{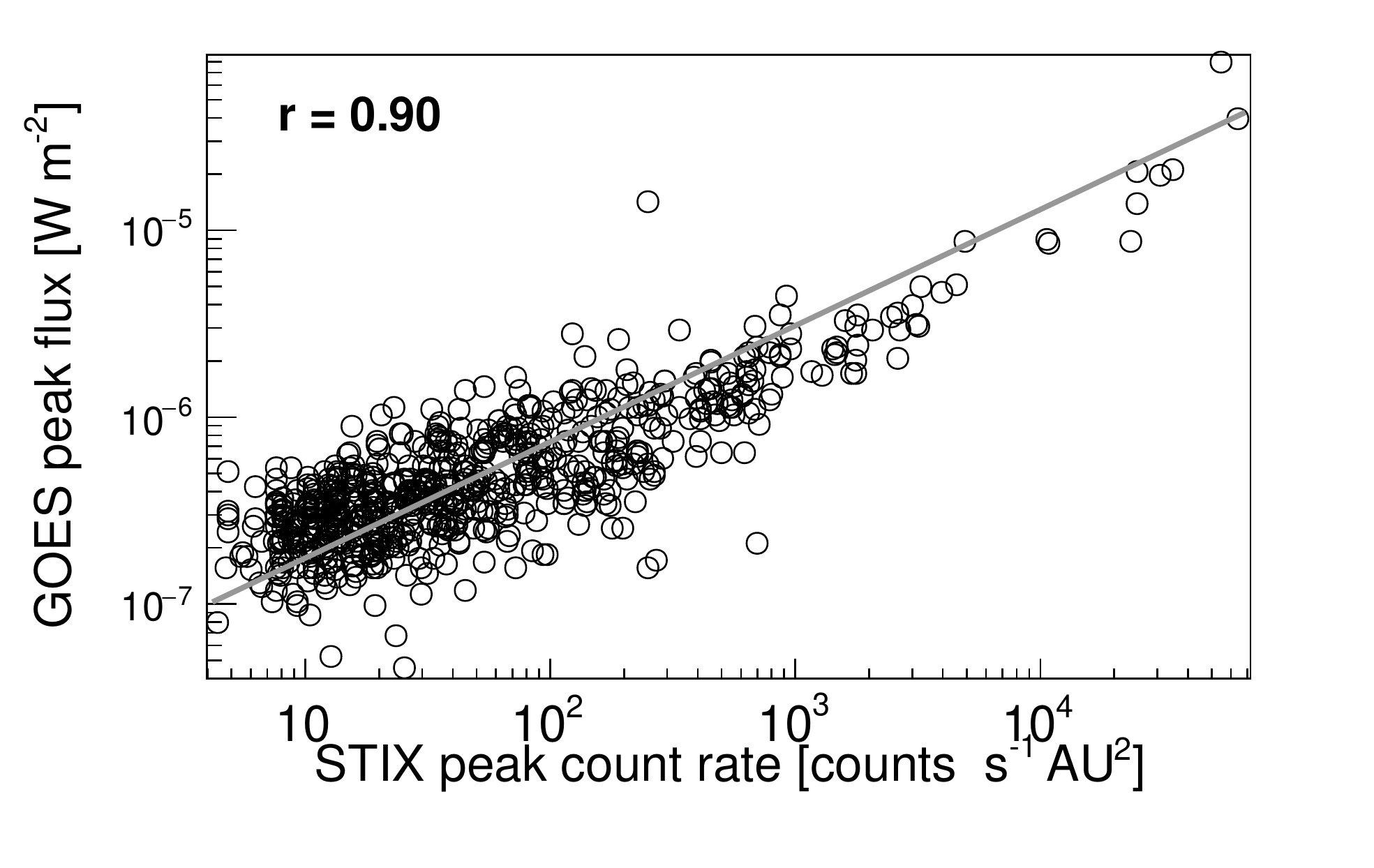}
  \caption{Scatter plot of GOES/XRS low channel peak flux with respect to STIX 4 -- 10 keV count rates scaled to 1~AU for 717 solar flares observed by both GOES and STIX during the cruise phase.   The solid line is a linear fit to the log-log graph.  The Pearson’s correlation coefficient  in the log-log scale is 0.91.   From the fit, we get the following GOES flux estimation formula: 
$f = 10^{-7.376+0.622 \log_{10} (X^{'})}$ (in units of W/m$^2$), where $X^{'}$ is the STIX peak count rate corrected for the distance variations between the Sun and Solar Orbiter. 
  }
\label{fig:goes-stix}
\end{figure}
Solar Orbiter is often far from the Sun-Earth line and hence a considerable number of flares observed by STIX 
are not observed by GOES satellites (and vice versa). 
In order to estimate the GOES classes of such flares,  we selected 717 solar flares observed by both GOES/XRS and STIX. 
Fig.~\ref{fig:goes-stix} shows the scatter plot of the peak fluxes
measured by GOES satellites with respect to the STIX background-subtracted count rates at the peaks, 
 in the energy range of 4 to 10 keV. 
STIX count rates $x$ have been corrected for 
the different distances between Solar Orbiter and the Sun using $X^{'}=x r^2$,
where $x$ is the count rate after background subtraction
 and $r$ is the distance between the Sun and Solar Orbiter in units of AU.  A clear correlation (The correlation coefficient $r=0.90$) is seen in Fig.~\ref{fig:goes-stix}.  The wide spread at low fluxes can be explained by the difference in 
the energy response of the two instruments and the variation in flare temperatures. 
The correlation can be fitted with a linear fit in the log-log scale. 
From the fit, we get the following GOES flux estimation formula: 
$f = 10^{-7.376+0.622 \log_{10} (X^{'})}$ (in units of W/m$^2$), where $X^{'}$ is the STIX peak count rate corrected for the distance variations between the Sun and Solar Orbiter. It is currently used to estimate the GOES classes of flares that are not directly observed by GOES satellites.  The estimated GOES fluxes are stored in the flare list collection in the database. 

\subsubsection{Estimation of coarse flare locations using CFL data}
STIX estimates approximate flare locations on board by 
maximizing the correlation between observed counts recorded by the 12 pixels in its coarse flare locator (CFL) subcollimator with expected counts using a lookup table \citep{stix2020}. 
With the requested science data, coarse flare locations can be reconstructed more accurately, as it allows for more sophisticated algorithms, as well as greater flexibility in selecting a time and energy range to be integrated. 

When the pixel data of a flare are available at the STIX Data Center, its flare location is estimated. 
The steps are as follows:
\begin{enumerate}
    \item Counts are integrated around the peak for each CFL pixel.
    \item The background is subtracted using a background file.
    \item The illuminated area is estimated on each CFL pixel.  This is based on two assumptions:  the illuminated area of a pixel is proportional to its relative count rate, and the total illuminated area of the imaging detectors is independent of the source location.
    \item The flare location is estimated by minimizing the weighted sum of squared deviations (i.e., weighted chi-squares) between the calculated illuminated areas and expectations simulated for potential flare locations in a 400 $\times$ 400 grid.
\end{enumerate}

\begin{figure*}
  \centering
  \includegraphics[width=0.95\linewidth]{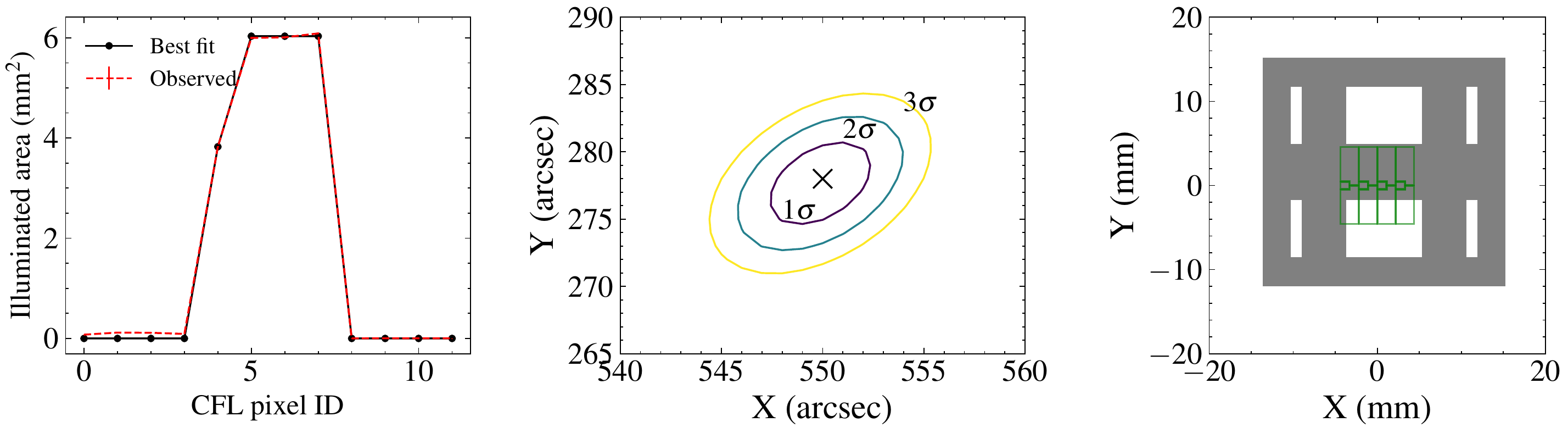}
  \caption{
Example of a  coarse flare location solution. 
	Left: Calculated areas of illuminated regions of the 12 CFL pixels and the best-fit simulated pattern for the flare location at (550, 278) arcsec. Pixels 0 to 3 are the top big pixels, pixels 4 to 7 are the bottom pixels, and pixels 8 to 11 are the small pixels as shown in the right panel.
   Middle: Best-fit flare centroid location (marked by x) and its 1$\sigma$, 2$\sigma$, and 3$\sigma$ confidence contours. {Right: Projection of CFL subcollimator (the gray shaded regions) on the detector pixels (the green closed areas) simulated for the best-fit flare location.  The top row,  bottom row, and small pixels (from left to right) are numbered sequentially from  0 to 11.  }}
  \label{fig:cfl}
\end{figure*}
As an example, the left panel of Fig.~\ref{fig:cfl} shows the calculated and best-fit illuminated areas of the CFL pixels for the flare observed at 2021-05-07T19:00:00 (GOES class M3.9);  the middle panel shows the best-fit flare centroid location, as well as its $1\sigma$, $2\sigma$, and $3\sigma$ contours. 
The simulated CFL shadow pattern is shown in the right panel. 
The estimated flare locations are stored in the flare list in the database. 

\subsubsection{Imaging and spectroscopy pipeline}

\begin{figure*}
  \centering
  \centering
     \begin{subfigure}[b]{0.495\textwidth}
         \includegraphics[width=0.95\textwidth]{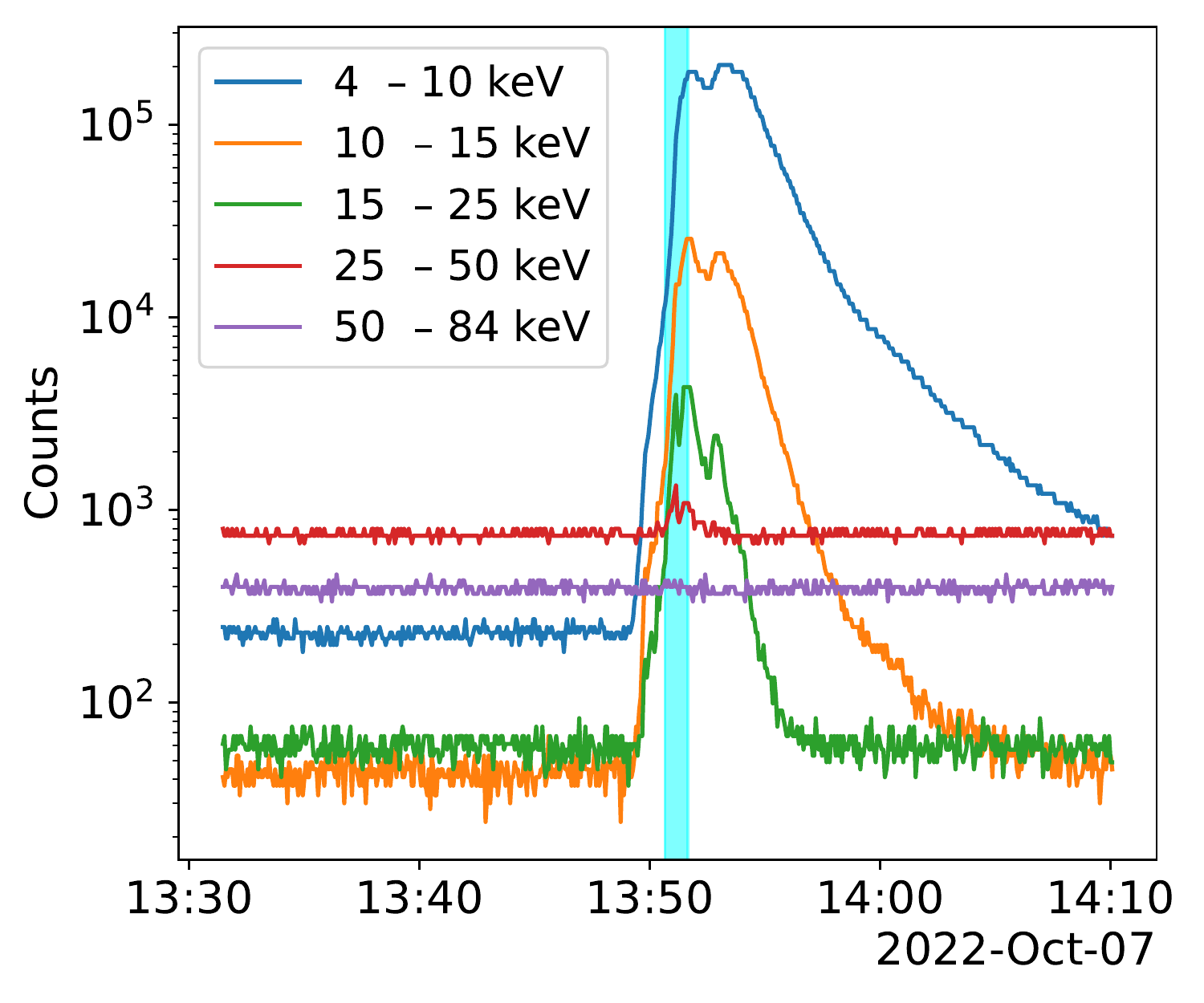}
    \end{subfigure}
     \begin{subfigure}[b]{0.495\textwidth}
         \includegraphics[width=1\textwidth]{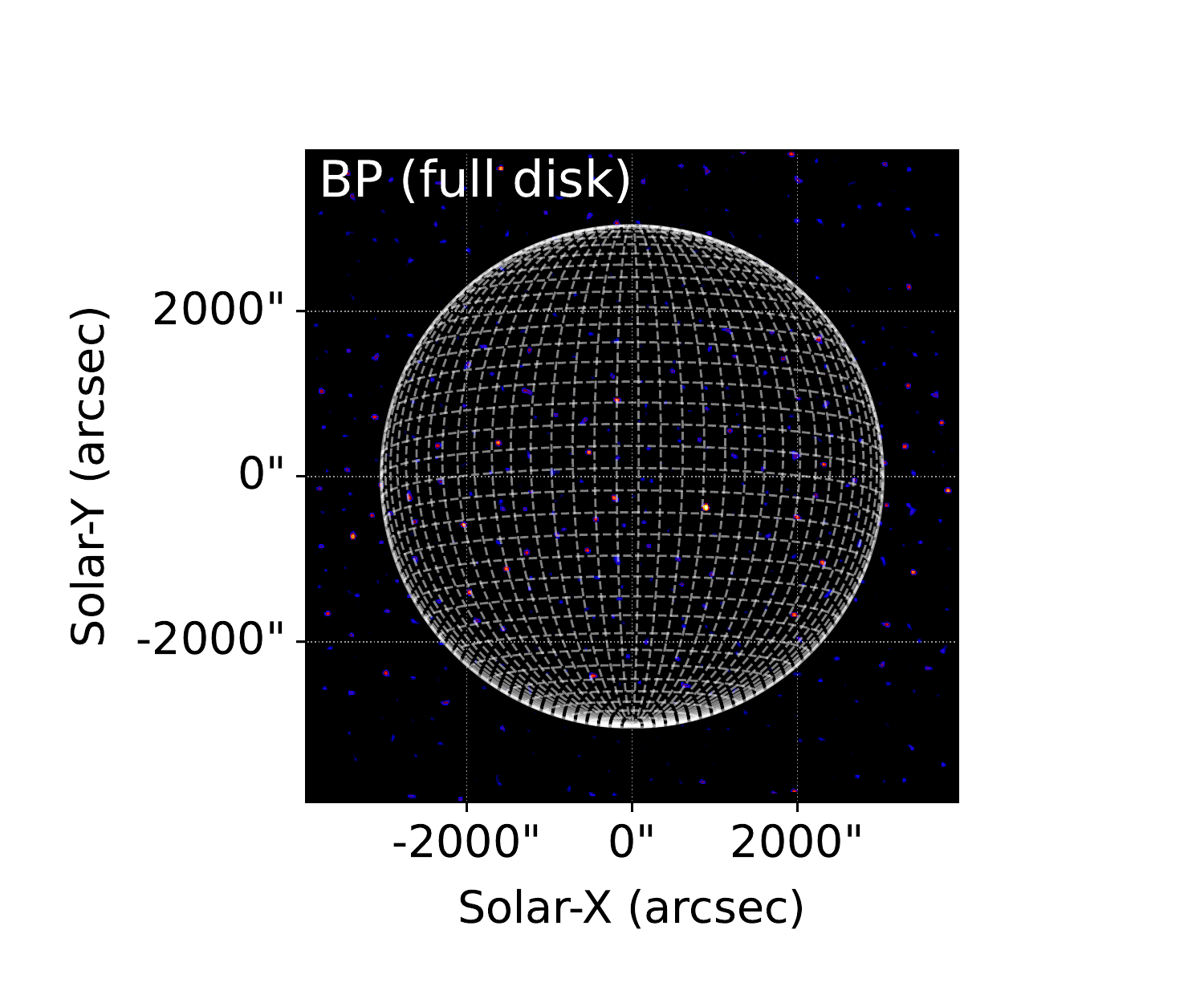}
    \end{subfigure}
     \begin{subfigure}[b]{0.495\textwidth}
         \includegraphics[width=1.05\textwidth]{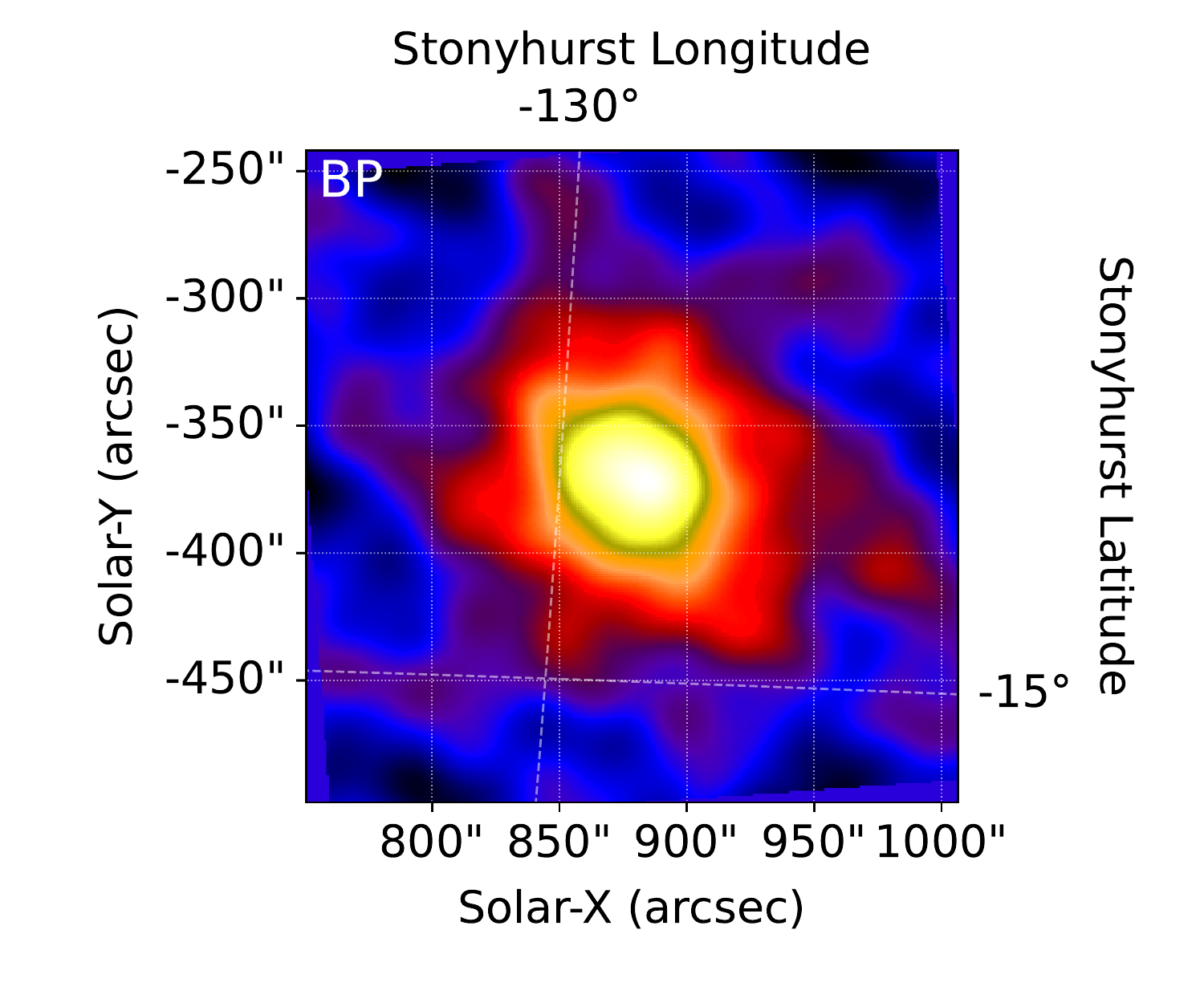}
    \end{subfigure}
     \begin{subfigure}[b]{0.495\textwidth}
         \includegraphics[width=1.05\textwidth]{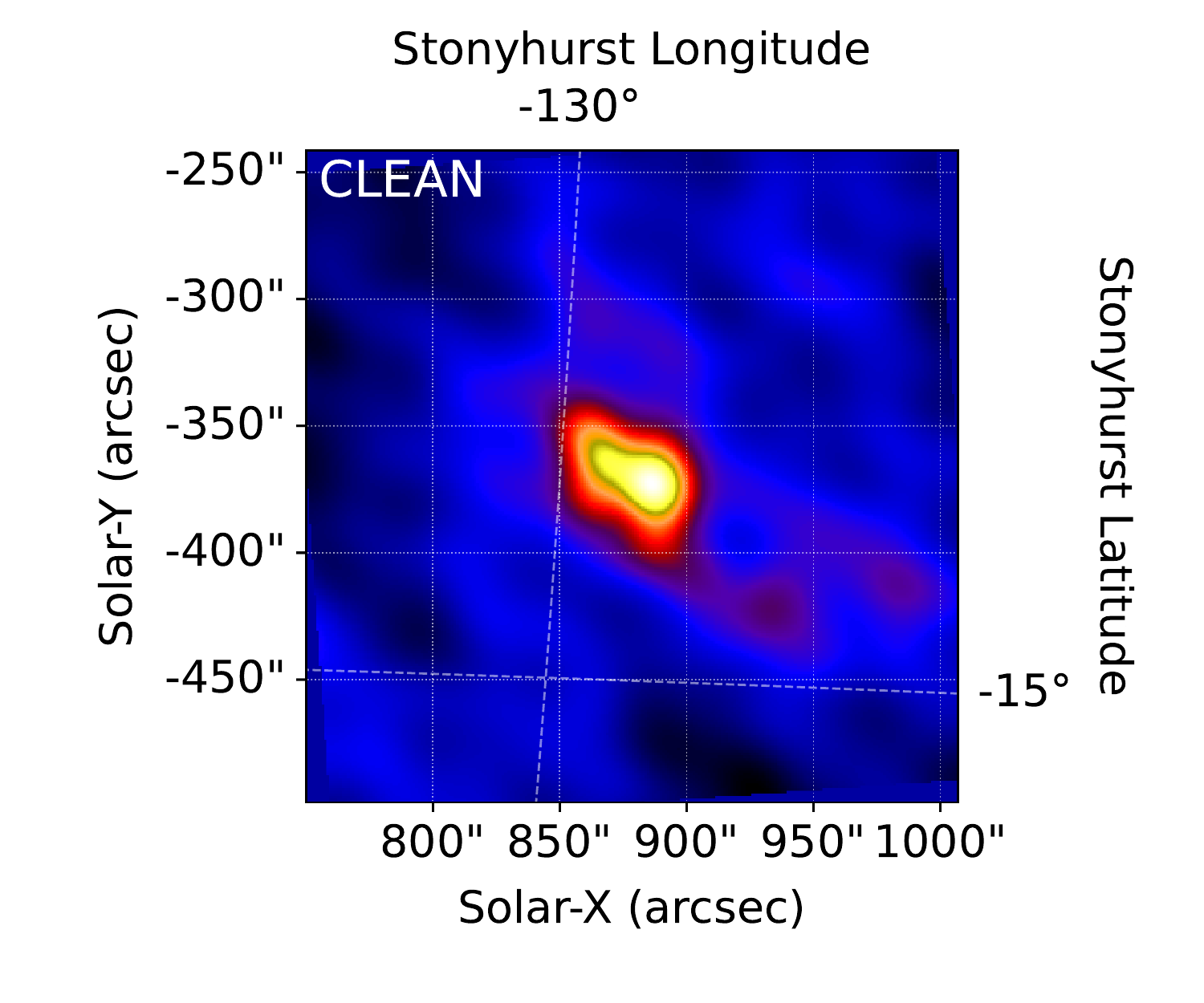}
    \end{subfigure}
     \begin{subfigure}[b]{0.495\textwidth}
         \includegraphics[width=1.05\textwidth]{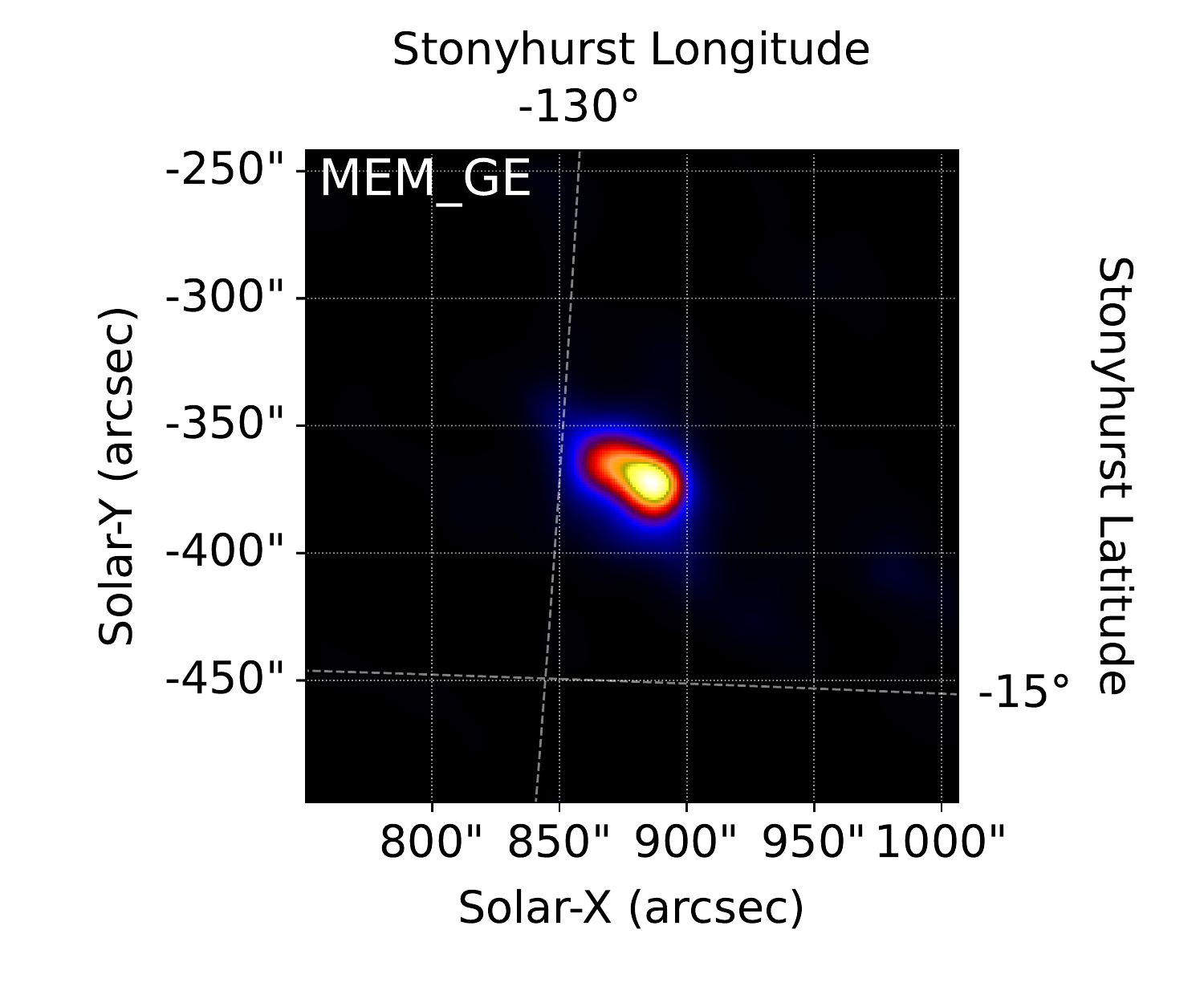}
    \end{subfigure}
     \begin{subfigure}[b]{0.495\textwidth}
         \includegraphics[width=1.05\textwidth]{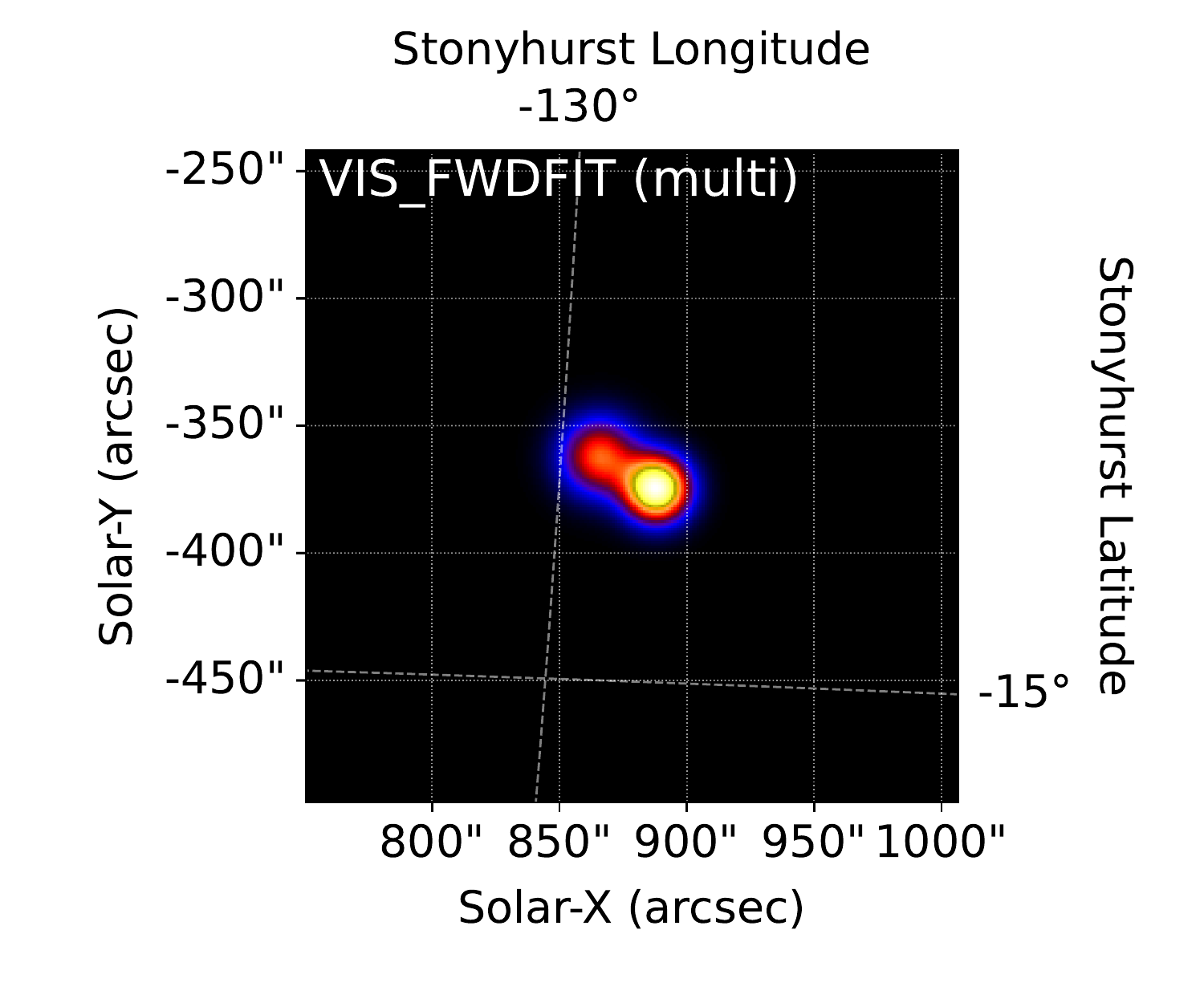}
    \end{subfigure}
    
  \caption{ 
   STIX QL light curves and reconstructed images of the solar flare observed at 
   2022-10-07T13:50:39 (UTC), 
   created by the image reconstruction pipeline.  A 1-min integration time around the peak was selected and 
     the images were reconstructed using four different algorithms: Back-Projection (BP), CLEAN, MEM\_GE, and VIS\_FWDFIT. 
The full-disk BP image shown in the top row of the right panel has been used to identify the source location. The other reconstructions were performed around the location of the flare. We note that the multiple circular Gaussian shapes have been selected for reconstructing the flaring source by means of VIS\_FWDFIT (bottom-right panel). }
  \label{fig:imaging}
\end{figure*}

\begin{figure}[h]
  \centering
  \includegraphics[width=0.95\linewidth]{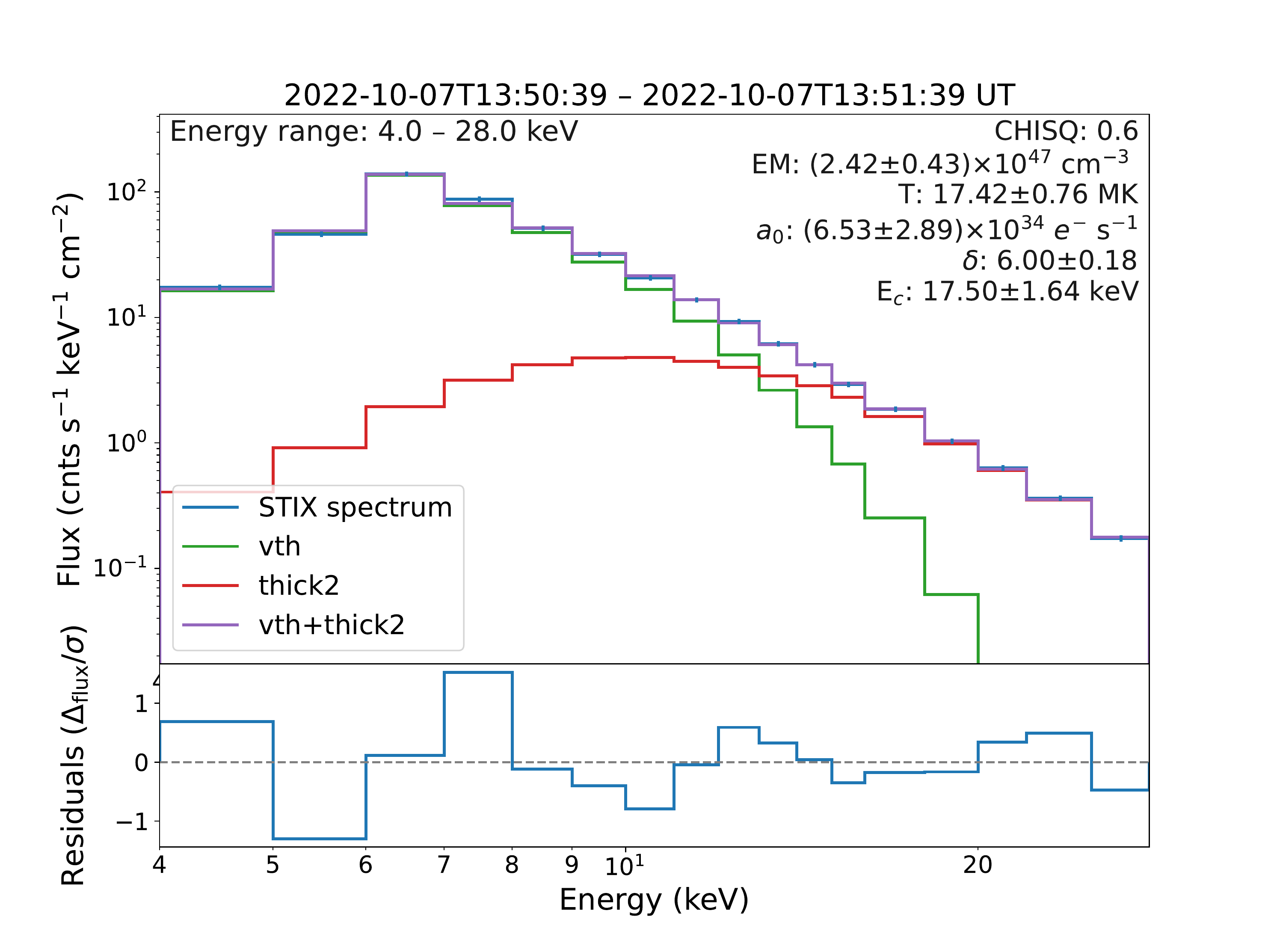} 
  \caption{Spectral fitting results for the same flare as in Fig.~\ref{fig:imaging}. The spectrum is fitted with a thermal component and a nonthermal component.}
  \label{fig:ospex}
\end{figure}

To help identify flares of interest, we developed an imaging and spectroscopy pipeline that automatically reconstructs images and performs spectral analysis for each flare with detector summed counts greater than 10,000
after receiving its pixel data. 
The pipeline first selects and integrates counts for 60 seconds around the peak of the flare for each pixel, and then subtracts the background from the integrated counts using the pixel data acquired during a quiet solar period. It is worth mentioning that quiet solar periods are determined on the basis of counts in the QL light curves. 
Subsequently, the transmission and dead-time corrections are performed on the background-subtracted counts. Then the corrected counts are further converted into the visibilities of two energy bands of 4 -- 10 keV (thermal energy) and 16 -- 28 keV (nonthermal energy). 
Then the visibilities are used for image reconstruction using four different algorithms: Back-Projection (BP) \citep{paolo2022}, CLEAN \citep{clean}, MEM\textunderscore GE \citep{mem, memge}, and VIS\textunderscore FWDFIT \citep{visfwd}. Finally, the reconstructed images are corrected for spacecraft off-pointing and rotations. 
As an example,  the first panel of Fig. ~\ref{fig:imaging} shows the light curves and time range selected for a flare that occurred at 2022-10-07T13:50:39 (UTC). 
The rest of the panels show the images reconstructed with the algorithms. 
In addition to image reconstruction, the transmission and the dead-time corrected counts are used for spectral analysis. Fig.~\ref{fig:ospex} shows the results of the spectral fitting for the same flare as in Fig.~\ref{fig:imaging}. The spectrum is fitted with a thermal component and a nonthermal component. 

The results of the pipeline are saved to files in both FITS and PNG formats. At the same time, the indexing information from the file, the parameter values of the spectral analysis, and the auxiliary data are recorded in a collection in the database. 

\section{Science data request strategy}
\label{sec:datareq}
As mentioned previously, high-resolution pixel data are only downlinked in response to requests from the
ground. They are stored in the onboard archive memory for a few months before being overwritten by new data. They are processed and downlinked 
after receiving data request telecommands from the ground. 
A data request telecommand contains information about the selected data and values of parameters  required to process the data on board, including the data compression level, 
time range, minimal time bin, energy bin width, 
and  masks indicating detectors and pixels to be selected. 
STIX detects thousands of flares per year; therefore, 
selecting data is a tedious task, as many factors must be considered,  such as count rates, time binning of data, statistics of selected data, and also the telemetry budget. 
The data selection strategy has been continuously optimized over the past two years. The current strategy is as follows: 
\begin{itemize}
  \item  
 Compressed L1 pixel data are requested for each of the ground-identified  flares with a total number of signal counts in the QL light curves greater than 10,000, which is approximately the minimum number of counts needed to reconstruct an X-ray image. 
The requested energy range is chosen to be the range in which obvious signal counts are seen,  
whereas, the requested time resolution is adjusted based on the telemetry budget and the scientific importance of the flare. 
If the peak count rate is above 125 counts/sec (approximately equivalent to the count rate observed for a B3 flare at 1 au), pixel counts with a high time resolution are requested.  Otherwise, pixel counts are integrated over the whole flaring time to reduce the telemetry data volume.   These parameters may be customized for individual flares by human operators during review.
 \item  Spectrograms with the highest time and energy resolution are requested for all periods when STIX is in its observation mode. 
 \item Time-integrated pixel data for background subtraction:
Time-integrated pixel data with durations of one or two detector temperature cycles (each cycle lasts about 40 minutes), which are acquired during quiet-Sun periods, are requested. The data are used for background subtraction when performing spectroscopy and imaging. 
\item High time-resolution aspect data are requested, for example, for periods when the spacecraft's attitude changes drastically.
Such periods can be known from the SPICE kernels or the aspect system readouts in HK data. 
\end{itemize}
The selection of science data of the above types is done automatically using a program (except for aspect data). The information of the selected data is written into a collection in the NoSQL database.  In addition, the STIX operations team also selects data for special 
needs.  After being checked and adjusted by the STIX operations team, groups of new data requests that meet the operations requirements are selected from the database and then compiled into instrument operation requests (IORs), which are used to create final telecommands at the MOC. 
These telecommands are then uploaded to the spacecraft and executed by STIX typically two to three weeks later. 

\begin{figure*}[ht]
  \centering
  \includegraphics[width=0.95\linewidth]{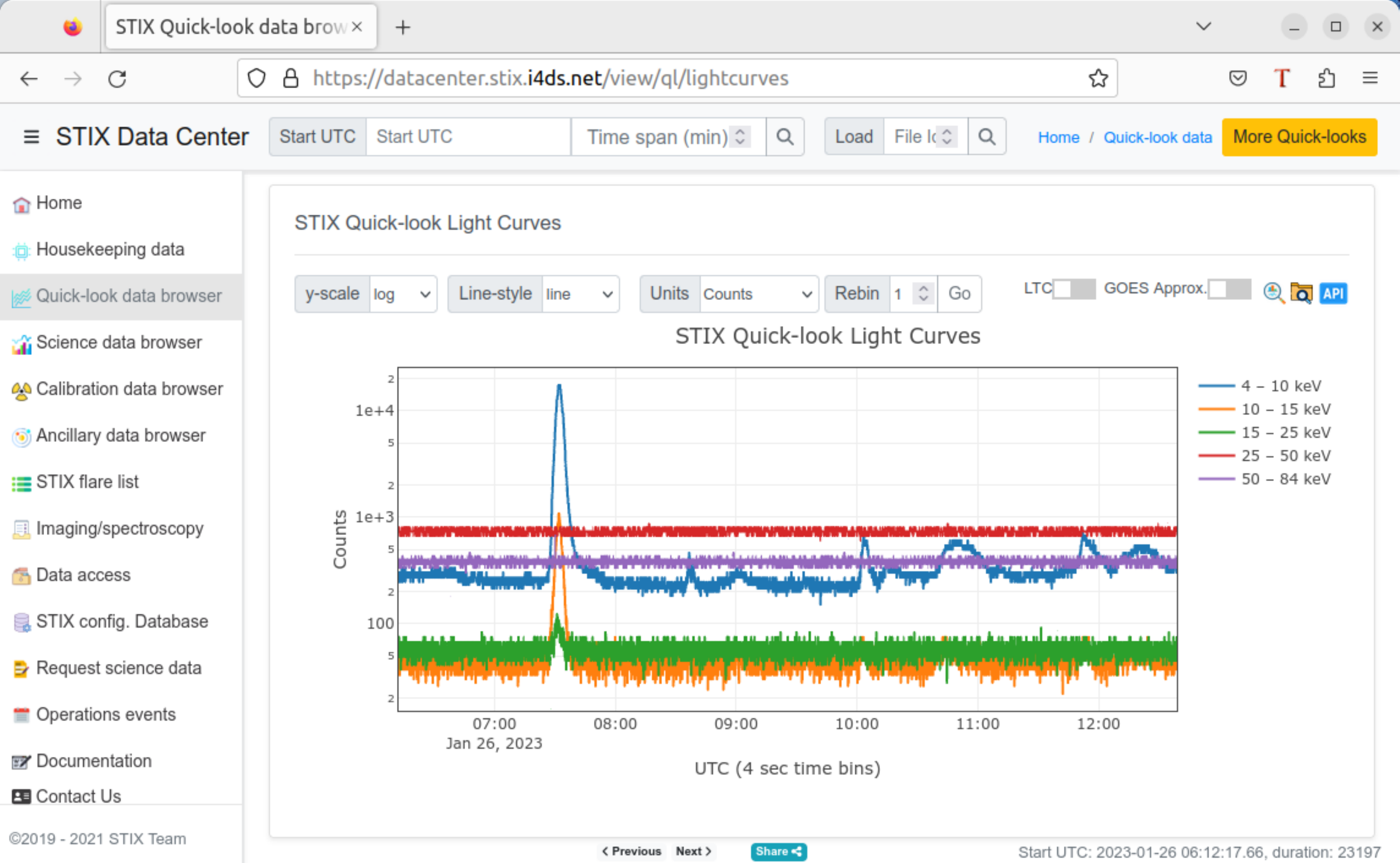}
  \caption{ 
    Interactive web-based STIX QL data browser. 
    In addition to STIX QLs, it can also display QLs of simultaneous measurements  performed by other instruments, such as GOES/XRS, SDO/AIA, and EUI on board Solar Orbiter.     The browser is available at the link \url{https://datacenter.stix.i4ds.net/view/ql/lightcurves}.}
  \label{fig:qlbrowser}
\end{figure*}
\begin{figure*}[ht]
  \centering
  \includegraphics[width=0.95\linewidth]{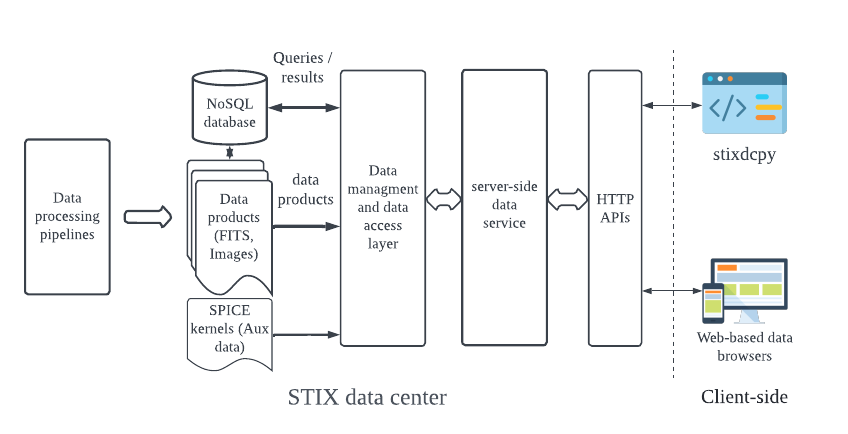}
  \caption{ 
   Client side and server side data exchange mechanism. 
  }
  \label{fig:interfaces}
\end{figure*}
\section{STIX Data Center user interfaces}
\subsection{Interactive web pages}

The data center platform provides various HTTP interfaces (APIs) that allow access to STIX data products
and the NoSQL database via HTTP requests. 
We have built dozens of web applications to manage and browse STIX data based on these APIs. 
Web techniques are chosen because they offer many advantages, such as clear cross-platform 
usability, broad access through browsers, rapid development, and easy maintenance.

As an example, Fig.~\ref{fig:qlbrowser} shows a screenshot of the STIX QL data browsing 
tool. It allows users to browse available QL data interactively. It 
interacts with the server through APIs.
The data exchange mechanism between the server side and the client side is shown in  
Fig.~\ref{fig:interfaces}. 
After receiving a request from the client side, the server retrieves QL counts from the NoSQL database for the user-specified time range. 
After excluding duplicates and merging,  the server sends QL counts and metadata in JSON format back to the client. The data are then used to create interactive light-curve plots using JavaScript on the client side. 
The interactive plot uses state-of-the-art web technologies that enable users to perform a range of operations, such as rebinning the integration time, correcting the light travel time between the spacecraft and the Earth, and exporting plot data to a local file.
In addition, QLs from other solar-observing instruments can also be displayed on the same page after users' activation, 
making it easier to find events of interest for joint analysis.

Based on similar concepts, tens of web tools have been developed to browse other STIX data products. The four most commonly used tools are listed below: 
\begin{itemize}
  \item  {\bf The science data manager and browser} provides users with tools for searching, downloading, visualizing, and analyzing science data. The interactive analysis tools allow users to select data of interest for standard analysis tasks, such as background subtraction and energy rebinning,  without installing additional software. 
  The algorithms are implemented on the client side using JavaScript. In addition, users can submit imaging and spectroscopy tasks to the server and view the results on the same page. This reduces the barriers to exploring STIX data for new users and is convenient for experienced users.

  \item  {\bf The preview images and spectroscopy product viewer} is a web-based tool for managing and viewing the imaging and spectroscopy results. The viewer also provides tools for plotting the time evolution of emission measures and temperatures,  creating animations of X-ray images for the selected runs, generating IDL or Python templates that allow for the same results  to be reproduced on local machines, and so on. 
  \item {\bf The auxiliary data viewer} allows the user to view auxiliary data, such as the spacecraft locations, velocity, and attitude, using data derived from the SPICE kernel and  pointing solutions computed from aspect system data.
  The viewer also provides tools to calculate the looking angles of flares and the coordinates of solar limbs within the STIX field of view.
  \item  {\bf The HK data browser} enables users to view time series of all STIX HK parameters, including the temperature, voltage, operation mode, memory status, etc. It provides great convenience for the instrument operations team to monitor the instrument status. 
\item {\bf The STIX data access page} offers users a variety of tools to search and download STIX data products. 
It also provides links to web tools to preview the products. As soon as they are generated at the STIX Data Center, STIX data products are immediately available for access on the page.
\end{itemize}

\subsection{STIX Data Center interface via  {\it stixdcpy}}
{\it stixdcpy} \footnote{stixdcpy source code is hosted on the GitHub repository at \url{https://github.com/i4Ds/stixdcpy}.} is a python package that facilitates accessing and analyzing STIX data. 
With {\it stixdcpy}, users can easily query and download the data products available at the STIX Data Center.
Similarly to the web tools, {\it stixdcpy} also provides tools to perform some standard analysis of STIX data, such as dead time correction, transmission correction, data clipping, and merging. 
{\it stixdcpy} is still under active development. 
As a result, its features and 
capabilities may change over time.  

\section{Summary}
\label{sec:summary}
STIX is one of ten instruments on board Solar Orbiter, 
which was launched into space on 10\ February 2020.
 STIX measures the  
intensity and spectrum of hard X-rays emitted during solar flares
in the energy range of 4 -- 150 keV.  
 During nominal operations, STIX continuously generates telemetry data. 
 To process and archive the data as well as to support the operations of the
 instrument and scientific activities using STIX data, 
 dedicated data-processing pipelines and a data platform have been 
 developed at the STIX Data Center.
 The pipelines generate telemetry at different levels and perform standard scientific analyses.  The data center platform distributes STIX data products and also provides users 
 with various web-based tools for searching and browsing STIX data products. 
  The data center is designed to work in a 
 fully automatic mode with minimal human intervention. The concept has proven successful 
 and has been running continuously for more than two years.
The data center not only facilitates the operations of the instrument, but it also provides great support to STIX data users.

\bibliographystyle{aa}
\bibliography{citations}

\end{document}